# Photostationary Lifshitz transition in High Tc superconductor $Bi_2Sr_2CaCu_2O_{8+\delta}$

Ji Dai[1,3†], Lukas Hellbrück[1,2†], Michele Puppin[1,2], Alberto Crepaldi[1,4], Francesco Barantani[1,2], Thomas LaGrange[1,2], Arnaud Magrez[1], Edoardo Martino[1], Neven Barisic[5,6], László Forró[1,7], J. Hugo Dil[1,2], Marco Grioni[1], Henrik M. Rønnow[1,2], Siham Benhabib[1,2,8*], Fabrizio Carbone[1,2*]

[1*]Institute of Physics, École Polytechnique Fédérale de Lausanne (EPFL), Lausanne, 1015, Switzerland.

[2] Lausanne Centre for Ultrafast Science (LACUS), École Polytechnique Fédérale de Lausanne (EPFL), Lausanne, 1015, Switzerland.

[3]LOREA, ALBA Synchrotron, Barcelona, 08290, Spain.

[4]Dipartimento di Fisica, Politecnico di Milano, Milan, 20131, Italy.

[5]Department of Physics, Faculty of Science, University of Zagreb, Croatia.

[6]Institute of Solid State Physics, TU Wien,Vienna, 1040, Austria.

[7]Department of Physics and Stavropoulos Center for Complex Quantum Matter, University of Notre Dame, Notre Dame, IN 46556, USA

[8]Laboratoire de Physique des Solides, Université Paris-Saclay, Orsay Cedex, 91405, France.
*Corresponding author(s). E-mail(s): siham.benhabib@universite-paris-saclay.fr; fabrizio.carbone@epfl.ch;
†These authors contributed equally to this work.

**Abstract**

To date, controlling the steady-state electronic band structure in high-Tc cuprate superconductors has been achieved primarily through chemical doping or magnetic fields. Here, we present that ultrafast optical excitation can instead drive the electronic band structure of $Bi_2Sr_2CaCu_2O_{8+\delta}$ into a photostationary, long-lived excited state. At sufficiently high pump fluences, this state undergoes a Lifshitz transition of the Fermi surface, characterized by a change in topology from hole-like to electron-like. Time- and angle-resolved photoemission spectroscopy, supported by single-band tight-binding calculations, reveals that 1.6 eV photoexcitation induces band-structure evolutions closely analogous to those produced by chemical doping. These results point to an efficient photodoping mechanism involving cooperative effects, including charge transfer, renormalization of effective electronic correlations, and defect-assisted charge trapping. Our findings raise fundamental questions regarding thermalization processes occurring on timescales comparable to the laser repetition period in cuprates. More broadly, ultrafast optical control enables access to otherwise inaccessible regions of the phase diagram by tuning the pump fluence.

## Introduction

The precise control of the quantum properties via ultrafast light pulses has opened avenues for exploring novel states without equilibrium analogs, so-called transient states. These transient states exhibit temporal changes in the free-energy landscape resulting from the reallocation of the charge carriers' populations, the dynamical renormalization of the coupling strengths, and coherent effects produced by resonant excitations of collective modes (*1*). Examples of such light-induced states include Floquet-Bloch states in topological systems (*2*), Insulator-to-Metal transition (*3*, *4*), switchable hidden quantum states (*5*, *6*), light-induced superconductivity (*7*, *8*), and coherent oscillations of the condensate in superconductors (*9*, *10*).

Over the past two decades, time-resolved studies have largely focused on ultrafast dynamics, often assuming that excitation relaxation occurs on timescales shorter than the laser repetition rate (*11*). Although most excited states relax within a few picoseconds, a small fraction persists much longer, exhibiting no clear evidence of complete relaxation between pulses. With increasing excitation fluence, these nonthermal states may become increasingly long-lived, extending to timescales far exceeding the laser repetition period. Such photostationary states are of particular interest in the context of nonthermal phase transitions, emerging from a reshaping of the free energy landscape towards a new global minimum, possibly stabilizing metastable hidden phases. These phases emerge from modifications to the free-energy landscape, which becomes sufficiently reshaped to form a long-lived local minimum that effectively replaces the global minimum, thereby enabling the stabilization of metastable hidden phases (*1*, *11*, *12*). A possible mechanism for the formation of a photostationary state is cumulative charge trapping at defect sites driven by repetitive photoexcitation. In this scenario, upon excitation with a train of laser pulses, a fraction of the generated photocarriers fail to recombine within the pulse period, leading to a progressive accumulation and a gradual modification of the electronic landscape. The modification reaches a steady-state, or saturation regime, once a dynamic balance between carrier injection and defect relaxation is established, the system reaches a long-lived stationary state, see Fig. 1B. When the optical excitation ceases, the trapped carriers are eventually released, allowing the system to relax back to its equilibrium ground state. It is important to note that photostationary states are not exceptional, as is often assumed. A recent review of ultrafast phenomena revealed that a significant amount of the investigated systems host a photostationary state (*11*). The misconception may arise from the fact that the detection of these states is not straightforward when using conventional stroboscopic methods. In fact, in most studies, the pump-induced effects are inferred by comparison with a reference taken at delays comparable with the laser repetition rate, and thereby not reflecting the true ground state in the presence of photostationary effects, see Fig. 1C.

This framework becomes particularly crucial in cuprates, where small changes of chemical doping, temperature, and magnetic field can lead to dramatically different ground states.

In hole-doped cuprates, a fine-tuning of the chemical doping leads to quasi-adiabatic modifications of the ground state, generating significant changes in the topology of the Fermi surface (FS). On the underdoped side, small hole-like Fermi arcs are located around the nodal regions along with (π, π) direction (*13*, *14*). As the doping increases, the FS transforms into a single large cylindrical, hole-like FS centered at (π, π) (*15*, *16*). At (0, π) direction, the band structure exhibits a flat dispersion, often referred to as the extended saddle point (*17*). This saddle point is a common feature of all cuprates, and in hole-doped cuprates, it appears near the Fermi level (*18*), whereas in electron-doped cuprates, it resides approximately 300 meV below the Fermi level (*19*). As doping increases in several cuprate families, such as $Bi_2Sr_2CaCu_2O_{8+\delta}$, $Bi_2Sr_2CuO_{6+\delta}$, $La_{2-x}Sr_xCuO_4$, and Nd-LSCO, the extended saddle point progressively moves toward the Fermi level. Upon crossing it, the FS undergoes an abrupt change in topology from hole-like to electron-like (*20–23*) see Fig. 1A, marking a Lifshitz transition (*24*). Interestingly, this Lifshitz transition is frequently accompanied by the disappearance of the pseudogap state (*21*), wherein the distinctive characteristics of the strange metal become more pronounced, signaling the presence of a quantum critical point (*25–27*).

In this work, we employ time- and angle-resolved photoemission spectroscopy (tr-ARPES) to demonstrate that ultrafast 1.6 eV excitation of optimally doped $Bi_2Sr_2CaCu_2O_{8+\delta}$ (Bi2212) stabilizes a photostationary state that, at sufficiently high fluence, undergoes a Lifshitz transition of the antibonding FS sheet from hole-like to electron-like. Our data reveal a pronounced photoinduced shift of the quasiparticle Fermi wave vector ($k_F$) across momentum space, with the strongest effect near the antinodal (0, π) regions. Single-band tight-binding simulations indicate that this shift reflects the motion of the extended saddle point toward the Fermi level with increasing excitation energy density, in close analogy to chemical doping.

**Results**

To identify the photostationary state in Bi2212, we performed tr-ARPES measurements at a fixed delay time of Δt~167 μs this was achieved by measuring the system at negative pump-probe delay times, 1 ps before the arrival of the pump pulse in experiments carried out at 6 kHz, see Fig. 1C. We also report the ultrafast dynamics for comparison with previous tr-ARPES studies in Bi2212; these data are presented in the Supplementary Information (SI).

Fig. 2 presents our main results. In Fig. 2A, we show the fluence dependence of the FS maps under 1.6 eV pump excitation, measured at a delay time of Δt~167 μs. We observe a progressive change in the FS topology as the pump fluence increases. At low fluence (F = 0.16 $mJ.cm^{-2}$), the FS closes around (π, π) and exhibits a hole-like character. As the fluence is increased to 1.1 $mJ.cm^{-2}$, it becomes electron-like, closed around (0, 0). Upon further increasing the fluence to 1.74 $mJ.cm^{-2}$, the FS continues to evolve, acquiring an increasingly electron-like character. The observed modification of the FS topology under ultrafast infrared excitation at this delay time is consistent with the formation of a photostationary state, marked by a gradual evolution from hole-like to electron-like topology at high fluences. It appears that once the laser pulses excite Bi2212, a stationary modification of the band structure is established, effectively becoming the system's new ground state. It is important to note that, in our measurements, the relaxed equilibrium electronic band is fully recovered within 60 s, as determined by the time required to acquire a new electronic band with sufficient statistics immediately after stopping the infrared pump. While the long lifetime makes it difficult to determine the exact timescale, we can establish a lower bound of 166.67 μs and an upper bound of 60 s for the duration of the effect. Nevertheless, Fig. 2A clearly reveals a high-fluence laser-induced photostationary modification of the FS topology, while finite energy and momentum resolution prevent a quantitative determination of the FS area and the corresponding carrier concentration. To overcome these limitations, we performed fluence-dependent single-spectrum scans, dispersions, along $k_{//}$ from the nodal to near-antinodal regions of the Brillouin zone, as shown in Fig. 2B; the full data set is presented in the SI.

The fluence range spans 0.09 to 1.15 $mJ.cm^{-2}$, and the electronic dispersions in Bi2212 was measured at two momenta, $k_{//}$=-0.410 $Å^{-1}$ and $k_{//}$=-0.535 $Å^{-1}$, at a fixed delay time of Δt~167 μs following 1.6 eV optical excitation. Additional measurements at two other momenta, $k_{//}$=-0.452 $Å^{-1}$ and $k_{//}$=-0.494 $Å^{-1}$, are presented in detail in the SI. Notably, we cannot reliably extract the dispersion at momenta higher than $k_{//}$=-0.535 $Å^{-1}$ due to broad incoherent QPs resulting from the pseudogap phase (*13*). Tracking the position of the QP peaks for increasing fluence reveals a notable shift in momentum towards 0 $Å^{-1}$ for the region close to the anti-node at $k_{//}$ = -0.535 $Å^{-1}$, as indicated by red arrows in Fig. 2B. Contrary, almost no changes were observed near the node at $k_{//}$= -0.410 $Å^{-1}$. To quantify the photoinduced $k_F$ shift, we extracted momentum distribution curves (MDCs) in a range of ±100 meV around $E_F$ and fit the QP peak position with Lorentzians for each fluence at the four $k_{//}$ cuts, as presented in Fig. 2B, with the fit results for all points summarized in Fig. 2C. It is noteworthy that the MDC peaks appear relatively broad and, in some cases, exhibit a multi-peak structure. We attribute this behavior to a possible non-uniform doping distribution within the probed region. To account for this inhomogeneity, the MDCs were fitted using a Voigt function, with the corresponding results presented in the Supporting Information. Importantly, fits obtained using both Voigt and Lorentzian functions show consistent trends, confirming the robustness of our conclusions regarding the fluence dependence.

At $k_{//}$ = -0.535 $Å^{-1}$, the difference in parallel momentum between the two sides of the band, left and right, decreases considerably with increasing fluence, resulting in a total reduction of 20% between fluences of 0.09 mJ.cm$^{-2}$ to 1.15 mJ.cm$^{-2}$. As we approach the node, specifically for $k_{//}$=-0.494 $˚A^{-1}$, $k_{//}$=-0.452 $˚A^{-1}$, and $k_{//}$=-0.41 $˚A^{-1}$, the difference in parallel momentum remains noticeable, although diminished. We observe a reduction of 5%, 3%, and 2.7% between the fluences of 0.09 mJ.cm$^{-2}$ to 1.15 mJ.cm$^{-2}$, respectively. Whereas a previous study reported such $k_F$ shifts only near the nodal regions and at low fluences up to 0.3 mJ.cm$^{-2}$(*28*), here we employ higher-energy high-harmonic-generation (HHG) probe pulses and substantially higher pump fluences in Bi2212, enabling the observation of a photostationary Lifshitz transition.

We emphasize that the QP $k_F$ shift is consistently observed across measurements on multiple samples; however, the magnitude of the photoinduced $k_F$ shift varies significantly from sample to sample (see SI). We attribute this variation to differences in sample quality, as the number of defects is expected to strongly influence the formation of the photostationary state through charge-trapping processes. Additionally, we carefully analyze the Fermi level across all fluences and observe that the Fermi level stays constant below a fluence of 1.8 mJ.cm$^{-2}$, ensuring that observed band structure modifications are not caused by space-charge artifacts, which would instead result in a shift and broadening to the entire band structure, including the Fermi level (*29*).

To elucidate the microscopic origin of the photostationary state, we performed an extensive modeling of the different cuts of the sample's FS under the different probed conditions. In the following, we concentrate on reconstructing the FS for the full pump fluence dependence, based on the four previously shown fluence-dependent $k_F$ shifts presented in Fig. 2C. The resulting $k_F$ points are then plotted in Fig. 3A for all fluences. The 4-fold symmetry of the crystal was used to better visualize the changes across the whole first BZ. Finally, we reconstructed the complete FS using a one-band tight-binding model (*30*), where μ, t, t′, and t′′ represent the chemical potential and the in-plane hopping parameters (Equation 1). The parameters and their corresponding values are provided in the SI.

$$\epsilon_k = \mu + \frac{t}{2}\left(\cos(k_x a) + \cos(k_y a)\right) + t'\cos(k_x a)\cos(k_y a) + \frac{t"}{2}\left(\cos(2k_x a) + \cos(2k_y a)\right) \pm \frac{t_{bi}}{2}\left(\cos(k_x a) - \cos(k_y a)\right)$$

In the bilayer cuprate Bi2212, the presence of two $CuO_2$ planes along with interlayer hopping causes the formation of split bonding and antibonding bands (*31*). The effect of the bilayer splitting is recognized in the single-band tight-binding model by the presence of the interlayer hopping parameter $t_{bi}$. In Fig. 3A, we report the evolution with excitation fluence of the reconstructed antibonding FS. The bonding FS and the tight-binding parameters used in our model and their evolution with fluence are detailed in the SI. At low fluences, the antibonding FS displays a hole-like shape centered around (π, π). As the energy density increases, significant changes in curvature occur near the (0, π) directions. At higher fluences ranging from 0.88 $mJ.cm^{-2}$ to 1.15 $mJ.cm^{-2}$, our fits suggest that the antibonding FS undergoes a Lifshitz transition, shifting from a hole-like to an electron-like structure centered around Γ. This behavior is consistent with the fluence-dependent FS maps shown in Fig. 2A. In contrast, the bonding FS remains hole-like, even at the highest energy density of 1.15 $mJ.cm^{-2}$. concomitantly, the light-induced changes in the FS are accompanied by an expansion of its area as the energy density increases, see Fig. 3B. Using the Luttinger theorem (*32*), we estimate the number of charge carriers (n) which it enhances with fluence, see Fig. 3C. The charge carriers increase significantly with the fluence, leading to a number of up to n = 1.29 holes per Copper site. To date, chemical doping through oxygen insertion in Bi2212 single crystals has only reached a doping level of p = 0.24, with a critical temperature of Tc = 47 K (*33*), indicating that light has a greater potential to achieve higher hole-doping levels. It is worth mentioning that our calculations do not consider the impact of the pseudogap, which partially suppresses electronic states near the antinodes, causing localization and a shift from n=1 + p to n=p (*34–36*), where p denotes the hole doping level.

## Discussion

The parent compounds of cuprates are Charge-Transfer insulators (*37*), where the strong Coulomb repulsion U between electrons in the Cu 3d orbitals generates a large energy gap separating the lower and upper Hubbard bands (LHB and UHB), with the O 2p orbitals lying in between the two, see Fig. 4. At finite doping, a metallic state appears inside the charge transfer gap ($\Delta_{CT}$), maintaining the reminiscence of the charge transfer phase (*38*). This new feature is generally attributed to a hybridized state between a local Cu 3d hole and a 2p hole on the four surrounding oxygens in the $CuO_2$ plane and is known as the Zhang-Rice singlet (ZRS) (*39*, *40*). These features indicate that, as in many strongly correlated systems, particularly transition-metal oxides, the dominant electronic interactions are governed by $\Delta_{CT}$ and Coulomb repulsion U. Consequently, any change in these interactions can lead to substantial modifications of the electronic properties. Illuminating cuprates with 1.6 eV ultrashort pulses has a significant impact on their electronic properties, predominantly by triggering charge-transfer ($\Delta_{CT}$) excitations, as this photon energy lies within the relevant energy scale. These excitations generally lead to a redistribution of charge among different orbitals within the structure. The dominant charge transfer occurs between the O 2p and Cu 3d orbitals (*41*, *42*), resulting in an enhanced population at the Fermi level and, consequently, an increase in itinerant carriers. This process appears to partially ungap the antinodal regions, as demonstrated by recent time-resolved ARPES studies (*42*). Furthermore, time-resolved X-ray Photoelectron Spectroscopy (tr-XPS) measurements have shown that 1.6 eV excitation induces not only charge redistribution between Cu-O atoms, but also involves the Bi-O reservoir layers and the Sr-O layers containing the apical oxygen (*43*).

Therefore, based on these observed effects of 1.6 eV excitation, we attribute the QP $k_F$ shift observed in our data to a photon-assisted doping process or photodoping, predominantly driven by $\Delta_{CT}$ excitations induced by 1.6 eV photons.

Moreover, additional effects arising from $\Delta_{CT}$ excitations must be considered, including changes to the effective Coulomb interaction. As presented above, triggering $\Delta_{CT}$ excitations with 1.6 eV photons leads to an increase in the FS area with fluence, reflecting an enhanced carrier density. Based on our Luttinger-theorem analysis, the carrier concentration increases from 1.17 to 1.29 holes per Cu site at a fluence of 1.15 $mJ.cm^{-2}$. This highly efficient carrier injection substantially enhances electronic screening, which in turn induces a significant renormalization of the effective Coulomb interaction. Consistently, such a renormalization of the on-site Coulomb repulsion at Cu sites induced by 1.6 eV photoexcitation has recently been observed using time-resolved X-ray absorption spectroscopy (tr-XAS) in underdoped $La_{1.905}Ba_{0.095}CuO_4$ under high excitation densities (*44*). The renormalization of effective electronic correlations by ultrafast pulses is not unique to cuprates; it has also been observed in the type-II Weyl semimetal Td-$MoTe_2$, where it is accompanied by a Lifshitz transition (*45*).

Furthermore, ultrafast pulses with high energy density have been shown to induce a transient structural transition signaled by the presence of an isobestic point in the temporal evolution of the electron diffraction signal from an optimally doped cuprate thin film (*46*), as well as to promote a highly anisotropic electron-phonon coupling in a Bi2212 single crystal (*47*). In addition, these ultrafast electron dynamics studies on both Bi2212 and $La_2CuO_{4+\delta}$ revealed an expansion of 2.5% in the size of the c-axis unit cell without any alteration of the crystal symmetry, triggered by 1.6 eV pulses (*46*, *47*). Such structural changes would significantly affect the vertical hopping parameter $t_{bi}$ in Bi2212, altering the topology of the FS.

The pressing question now is how to reconcile the dynamics of these cooperative effects, $\Delta_{CT}$ excitations, renormalization of the effective Coulomb interaction, and transient structural changes, that evolve on ultrafast timescales, ranging from hundreds of femtoseconds to hundreds of picoseconds, and collectively generate efficient photodoping that persists up to microseconds in the form of a stabilized photostationary state.

A possible scenario is that, following the arrival of the first pump pulse, a sequence of cooperative effects is initiated. $\Delta_{CT}$ excitations are triggered first, accompanied by a renormalization of the effective Coulomb interaction, and subsequently by transient structural changes. The majority of the electronic excitations thermalize on a timescale of 135-165 fs (*48*, *49*), followed by the relaxation of the effective Coulomb renormalization on a timescale of 220 ± 60 fs (*44*). However, several measurements, including our stroboscopic data as well as other time-resolved ARPES and time-resolved optical spectroscopy studies (*48–50*), see SI, reveal an absence of complete thermalization even at long time scales, extending up to several tens of picoseconds. While most excited states relax within a few hundred femtoseconds, a minority persist much longer, showing no clear evidence of complete relaxation within the laser repetition period. Notably, as the excitation fluence increases, the persistence of these non-thermal states becomes more pronounced and extends over longer timescales, as demonstrated by transient reflectivity measurements (*50*). In the presence of structural defects, particularly vacancies, a minority of the excited charges become trapped at these sites. In Bi2212, the main source of defects is oxygen dopants, which have been shown through shot-noise measurements combined with STM to be localized between the Bi-O and Sr-O layers (*51*, *52*). These dopants give rise to states at energy scales of -1eV and -1.5 eV (*53*). The number of trapped charges is proportional to the nature and concentration of vacancies and depends on their spatial distribution, all of which are intrinsically linked to the synthesis conditions, explaining why our measurements across multiple samples reveal a strong sensitivity of the observed effects to sample quality. In Bi2212, the trapped minority charges do not relax back within the laser repetition period, which in our case is t~167μs, leading to a partially excited state. When the subsequent pump pulse arrives, the cycle repeats, and with each pulse, the partially excited state gradually builds up. This process evolves toward an optimal occupation of trapping vacancies governed by the balance between photocarrier injection and defect relaxation, leading to saturation and the stabilization of a long-lived excited state.

It is worth noting that early investigations of high-Tc superconductivity, employing static transport measurements, reported the existence of long-lived photoconductivity, along with the lesser-known phenomenon of an enhancement of the superconducting transition temperature Tc (*54*, *55*). It has been demonstrated that extended exposure to continuous light can alter the electrical properties of cuprates, frequently resulting in an enhancement of the superconducting transition temperature Tc in $YBa_2Cu_3O_y$ (YBCO). The proposed mechanism suggests that photoexcitation generates electron-hole pairs within the charge transfer gap. Electrons are subsequently trapped at vacancies and other defects, while the remaining holes populate the conduction band. This process effectively photodopes the material and increases the number of mobile charge carriers (*54*, *55*). Furthermore, recent combined transport and X-ray experiments on illuminated cuprates have linked the observed increase in photoconductivity to light-induced defect reordering (*56*). This is particularly relevant for Bismuth-based cuprates, which are known to host a high density of such defects, especially when increasing the hole-doping content through chemical substitution (*57*). A quantitative assessment of defects requires systematic resistivity measurements, with particular focus on the residual resistivity at low temperatures.

Driving the FS of cuprates through a Lifshitz transition has significant implications for the underlying electronic phases, particularly the pseudogap, as numerous experimental and theoretical studies have shown a strong connection between the onset of the pseudogap and the FS topology (*21*, *58*, *59*). In particular, it has been shown that the pseudogap, characterized by the presence of hot spots the regions with strong scattering and short lifetime, can only form when the FS is hole-like and intersects the antiferromagnetic zone boundary. Our results suggest that high-energy-density ultrafast light pulses can suppress the pseudogap by inducing a topological change in the FS, indicating that light can act as a novel control parameter, beyond chemical doping, capable of navigating the cuprate phase diagram. The persistent nature of the light-induced state makes excitation energy density a particularly attractive tuning parameter, enabling, for example, the exploration of quantum criticality by smoothly driving the system across a quantum critical point. More broadly, this approach offers a novel route to study the cuprate phase diagram in situ, using only excitation energy density and sample temperature as external control parameters. Moreover, the photostationary character of the system opens new experimental possibilities: transport measurements synchronized with laser excitation could allow direct investigation of the out-of-equilibrium photostationary state in the heavily hole-doped regime. Such studies may provide crucial insights into its microscopic origin and potentially reveal emergent exotic phases stabilized under nonequilibrium conditions.

## Materials and Methods

In this study, we investigated the transient electronic dispersion of slightly overdoped Bi2212 with a Tc of 89 K via ARPES using higher harmonic generation (HHG) lasers to produce extreme ultraviolet pulses at 27 eV and 30 eV at the Harmonium beamline within the LACUS facility at EPFL (*60*). The sample was cleaved in the measurement chamber in situ at a temperature of 77 K under an ultrahigh vacuum of $4\times10-11$ mbar. Photoexcitation was achieved using 780 nm ($\hbar\omega$ = 1.6 eV) femtosecond laser pulses at a repetition rate of 6 kHz, with a pulse duration of less than 50 fs, generated by a Ti: Sapphire amplified femtosecond laser. The relatively low repetition rate enabled excitation of the Bi2212 system with high fluence up to 3 mJ/cm2 without fast sample degradation. The overall energy resolution amounted to less than 200 meV and a time resolution of a 100 fs.

**Figures**

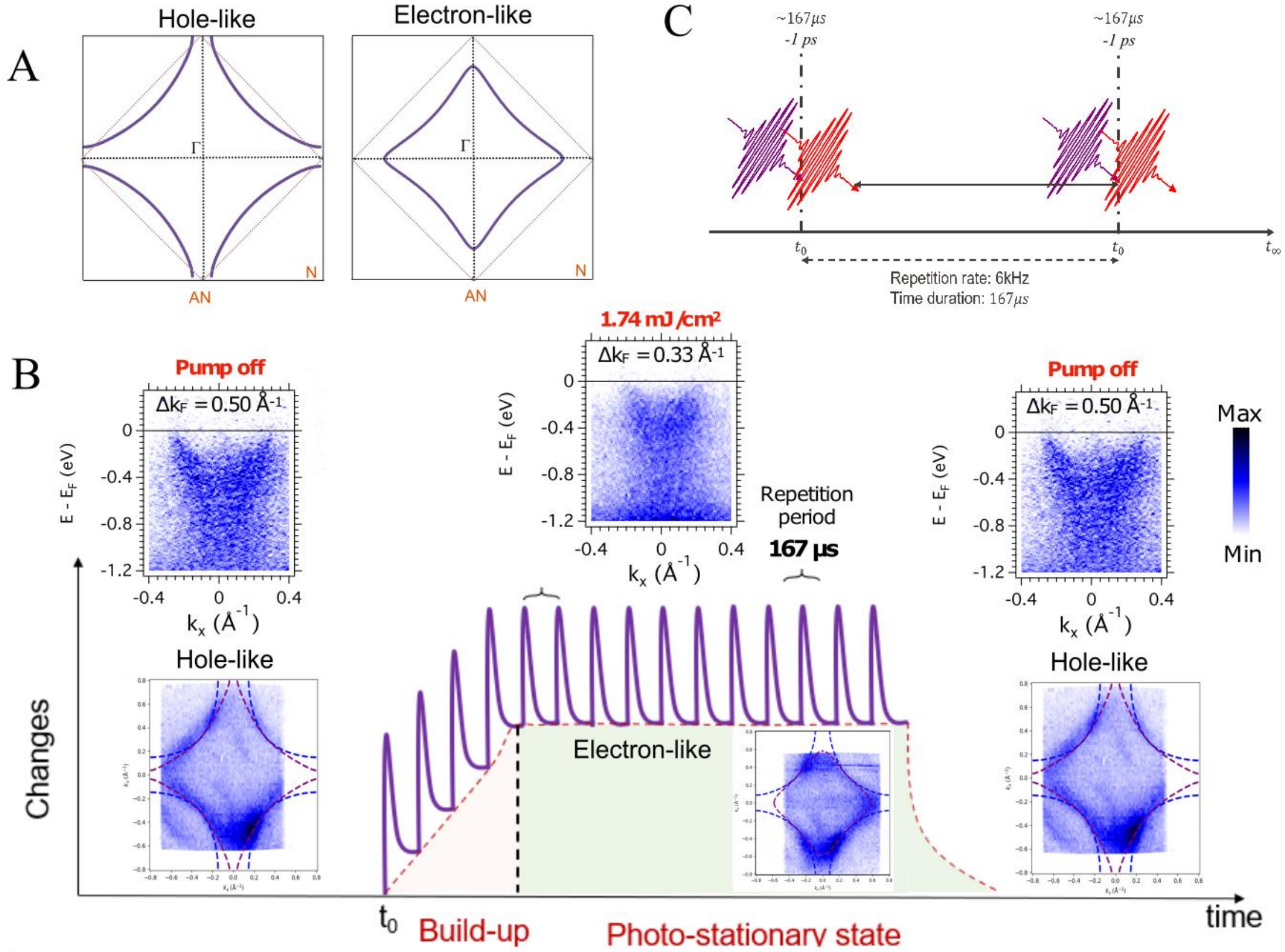

**Fig. 1. The photo-stationary states in Bi2212. (A)**: Schematic illustration of the evolution of the Fermi surface topology of Bi2212. The antibonding sheet is shown; the dashed square indicates the first antiferromagnetic Brillouin zone. N and AN denote the nodal and antinodal directions, respectively. **(B):** A schematic representation of the stabilization process of the photostationary state and the corresponding quasiparticle states in Bi2212 at $k_{//}$=-0.39 $Å^{-1}$, measured over a delay range of $\Delta$t ~167 μs at negative pump-probe delay times, 1 ps before the arrival of the pump pulse and the corresponding Fermi surface maps are shown for two conditions: pump off (zero fluence) and under excitation with a fluence of 1.74mJ.$cm^{-2}$. **(C)**: Schematic illustration of the fixed delay time used in the measurements $\Delta$t ~ 167 μs.

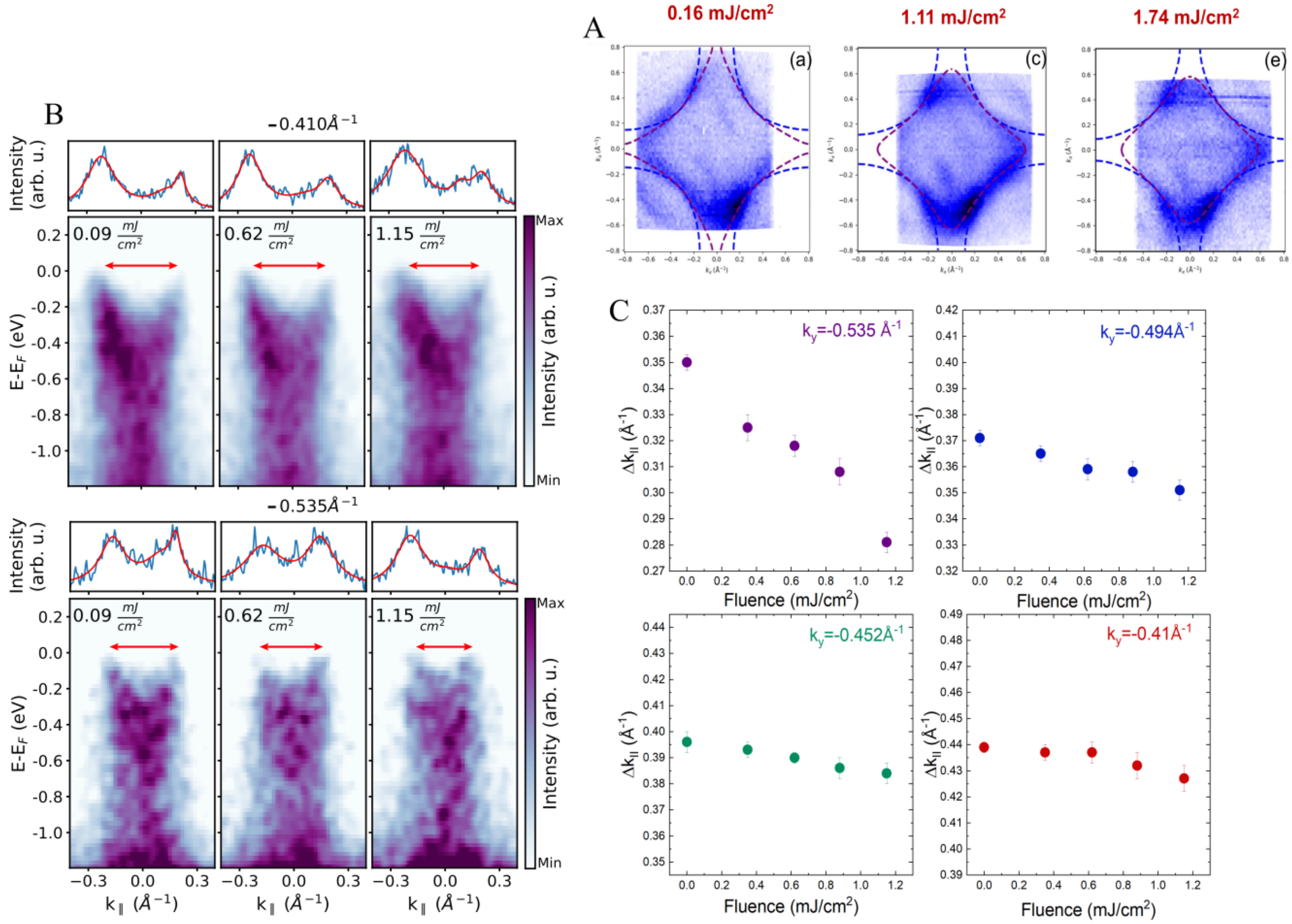

**Fig. 2. Pump fluence dependence**. **(A)**: Fermi surface maps measured at hν = 30 eV with pump fluences of 0.16 mJ cm$^{-2}$, 1.11 mJ cm$^{-2}$, and 1.74 mJ cm$^{-2}$, respectively, acquired at fixed delays of Δt ~ 167 μs. The blue and purple dashed lines indicate the fitted bonding and antibonding Fermi surfaces obtained from a single-band tight-binding model. **(B)**: Energy dispersion spectra of optimally doped Bi2212 at two $k_{//}$ points ($k_{//}$ = -0.535 Å$^{-1}$ and $k_{//}$ = -0.41 Å$^{-1}$) for pump fluences of 0.09 mJ cm$^{-2}$, 0.62 mJ cm$^{-2}$ and 1.15 mJ cm$^{-2}$respectively. MDCs for each dispersion are fitted with three Lorentzian functions representing the left and right QP peaks and the possible crossing of the replica band. Red arrows highlight changes in parallel momentum at the Fermi level. **(C)**: Extracted $\Delta k_F$ at the four measured $k_{//}$ ($k_{//}$ = -0.535 Å$^{-1}$, $k_{//}$ = -0.494 Å$^{-1}$, $k_{//}$ = -0.452 Å$^{-1}$and $k_{//}$ = -0.41 Å$^{-1}$) as a function of pump fluence.

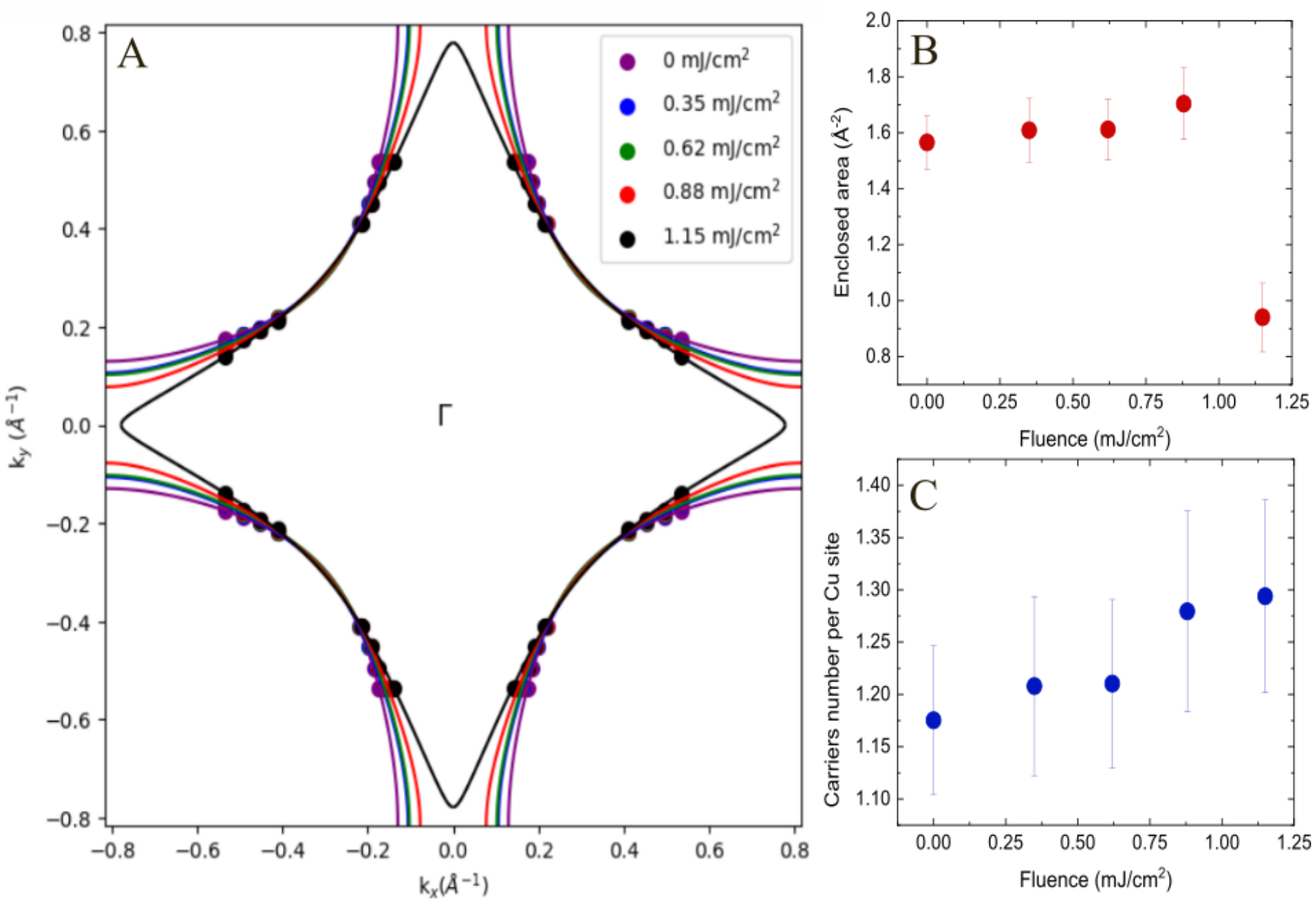

**Fig. 3. Fluence FS evolution**. **(A)**: experimental $k_F$ points at the Fermi level for the energy densities of 0.09 mJ cm$^{-2}$, 0.35 mJ cm$^{-2}$, 0.62 mJ cm$^{-2}$ 0.88 mJ cm$^{-2}$ and 1.15 mJ cm$^{-2}$ and the constructed one-band tight-binding Fermi surface antibonding sheet for all measured energy densities. **(B)**: The calculated enclosed area evolution with the energy density. **(C)**: The calculated number of charge carriers' evolution with the energy density

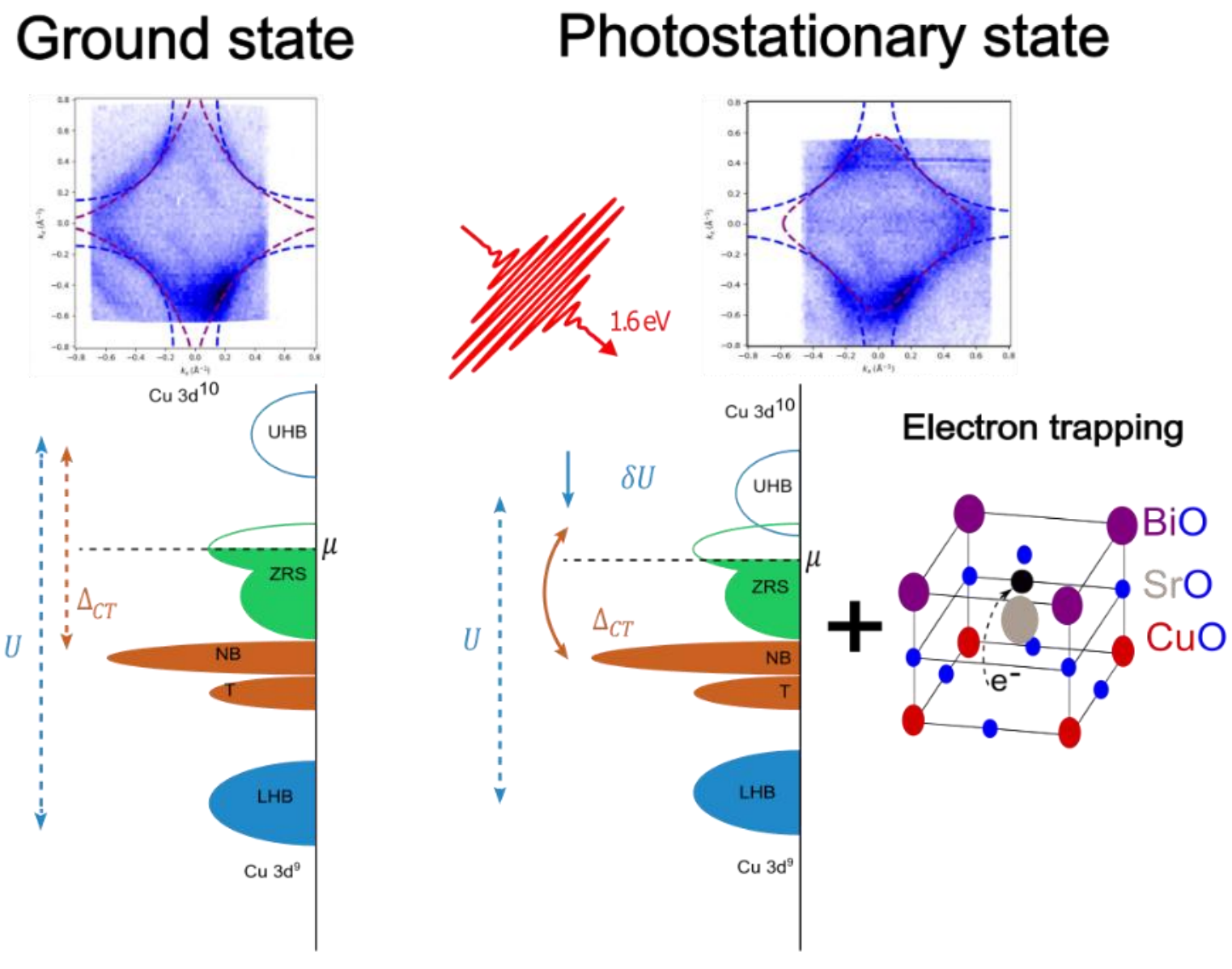

**Fig. 4. Schematic illustration of the mechanism of the formation of the photostationary Lifshitz transition in Bi2212. Left panel:** In the unpumped ground state, Bi2212

can be described within a Mott-Hubbard charge-transfer framework, with the Fermi level intersecting the Zhang–Rice singlet band and the electronic structure characterized by lower and upper Cu Hubbard bands separated by the on-site interaction $U$. **Right panel:** Upon 1.6 eV laser excitation, the photostationary state is described by a renormalized electronic structure, reflected in a modification $\delta U$ of the effective on-site interaction, together with CT excitations and carriers localized by trapping processes.